\documentclass[twocolumn]{aastex62}

\usepackage{natbib}
\usepackage{amsmath}
\usepackage{xcolor}

\definecolor{forestgreen}{RGB}{34,139,34}

\hypersetup{linkcolor=red,citecolor=blue,filecolor=cyan,urlcolor=magenta}

\graphicspath{{./}{figures/}}

\submitjournal{ApJ}

\shorttitle{Doppler Beaming of stars in the Galactic Center}
\shortauthors{M. Zaja\v{c}ek}

\begin{document}

\title{Enhanced Doppler Beaming for Dust-Enshrouded Objects and Pulsars in the Galactic Center}

\correspondingauthor{M. Zaja\v{c}ek}
\email{zajacek@cft.edu.pl}

\author[0000-0001-6450-1187]{Michal Zaja\v{c}ek}
\affil{Center for Theoretical Physics, Polish Academy of Sciences, Al. Lotników 32/46, 02-668 Warsaw, Poland}
\affil{Department of Theoretical Physics and Astrophysics, Faculty of Science, Masaryk University, Kotl\'a\v{r}sk\'a 2, 611 37 Brno, Czech Republic}
\affil{L. Novomeského 4, 901 01 Malacky, Slovakia}

\begin{abstract}

Stars within the innermost part of the Nuclear Star Cluster can reach orbital velocities up to a few percent of the light speed. As analyzed by Rafikov (2020), Doppler boosting of stellar light may be of relevance at the pericenter of stellar orbits, especially with the upcoming high-precision photometry in the near- and mid-infrared bands. Here we analyze the previously neglected effect of infrared spectral index of monitored objects on the Doppler-boosted continuum emission in a narrow band. In contrast to main-sequences stars, the detected compact infrared-excess dust-enshrouded objects have an enhanced Doppler-boosting effect by as much as an order of magnitude, with the variability amplitude of the order of ten percent for the most eccentric orbits. In a similar way, pulsars dominated by non-thermal synchrotron emission are also expected to exhibit a stronger Doppler-boosted signal by a factor of at least four in comparison with canonical S stars. In case the stellar orbit is robustly determined, the relative flux variation can thus provide hints about the nature of the objects. For extended dust-enshrouded objects, such as G1, that are variable due to tidal, ellipsoidal, bow-shock, and irradiation effects, the subtraction of the expected Doppler-boosting variations will help to better comprehend their internal physics. In addition, the relative flux variability due to higher-order relativistic effects is also modified for different negative spectral indices in a way that it can obtain both positive and negative values with the relative variability of the order of one percent. 


\end{abstract}


\section{Introduction} \label{sec:intro}

The Galactic center Nuclear Star Cluster (NSC) is a unique dynamical testbed, where it is possible to monitor the motion of stars in the potential of the compact radio source Sgr~A* \citep{1974ApJ...194..265B}, which is associated with the supermassive black hole of $\sim 4\times 10^6\,M_{\odot}$ \citep[SMBH;][]{2010RvMP...82.3121G,2014CQGra..31x4007S,2017FoPh...47..553E}. In particular, the innermost part of the NSC, the S cluster, contains several tens of bright B-type stars with inferred orbits \citep{2005ApJ...620..744G,2009ApJ...692.1075G,2017ApJ...837...30G,2020ApJ...896..100A}, which has been utilized for tests of gravitational theories in the strong-field regime \citep{2017PhRvL.118u1101H,2017ApJ...845...22P} as well as for $N$-body cluster dynamics in the potential dominated by the SMBH \citep[see e.g.][and references therein]{2013degn.book.....M}. 

For the S2 star on a highly elliptical orbit with the period of $\sim 16$ years, the combined general relativistic gravitational redshift and transverse Doppler shift of $\sim 200\,{\rm km\,s^{-1}}/c$ was measured using its spectral absorption lines \citep{2018A&A...615L..15G,2019Sci...365..664D}. The General Theory of Relativity was also confirmed to a high precision by the measurement of its Schwarzschild precession of $\delta \phi\sim 12'$ per orbital period \citep{2017ApJ...845...22P,2020A&A...636L...5G}. 


Recently, \citet{2020ApJ...905L..35R} analyzed the effect of nonrelativistic and relativistic Doppler boosting on the detected photometric flux density of S stars. The study was inspired by the analysis of the photometric lightcurve of S2 star in $L'$-band \citep{2020A&A...644A.105H}, who found no intrinsic variability at the level of $\sim 2.5\%$. \citet{2020ApJ...905L..35R} concluded that the Doppler boosting of the continuum stellar light is especially of relevance for stars that reach the pericenter velocities of a few percent of the light speed. The amplitude of the relative flux change due to the Doppler beaming is $\sim 2.0\%$ for the S2 star ($\Delta m\sim 0.02$ mag), and reaches $\sim 5.6\%$ for the newly analyzed S62 ($\Delta m\sim 0.06$ mag) and $\sim 6.4\%$ for the recently discovered S4714 star ($\Delta m\sim 0.07$ mag) \citep{2020ApJ...889...61P,2020ApJ...899...50P}. This is at the sensitivity limit of $\sim 0.01$ mag of current infrared adaptive optics instruments. The detection and the timing of the Doppler beaming for these sources is achievable with an order of magnitude improved sensitivity of $30+$ meter telescopes \citep{2019BAAS...51c.530D,2020A&A...644A.105H}. The higher-order corrections to the Doppler-beamed stellar light due to general relativistic redshift and transverse Doppler shift have the amplitude of the relative flux change at the level of $0.7\times 10^{-3}$ mag for S2, $\sim 2.6 \times 10^{-3}$ mag for S62, and $\sim 2.6\times 10^{-3}$ mag for S4714, which is close to the sensitivity limit of upcoming infrared instruments. However, they will hardly be detectable due to stellar crowding and confusion in the innermost parts of the NSC \citep{2020ApJ...905L..35R}. In principle, if the Doppler boosting effect and its temporal profile is detected, it could be utilized to verify and constrain orbital elements in case the photometry and the spectroscopy of the source is confused \citep{2020ApJ...905L..35R}.  

However, the overall uncertainty of the infrared spectral index of the Galactic center sources was not quantitatively analyzed by \citet{2020ApJ...905L..35R}. Since their study focused on B-type stars, such as S2, fixing the spectral index to $\alpha=2$ (using the convention for the radiation intensity as $I_{\nu}\propto \nu^{+\alpha}$) can be justified in the infrared domain. However, the S cluster also hosts several dust-enshrouded objects with a negative spectral index in the near-infrared domain due to the dust emission with the effective temperature of $\sim 500-600\,K$ \citep{2020A&A...634A..35P,2020Natur.577..337C}. In addition, the emission of compact remnants, in particular energetic pulsars, is expected to be dominated by optically thin synchrotron emission in infrared bands with $\alpha\sim -1$ \citep[see e.g.][and references therein]{2017A&A...602A.121Z}. Future surveys using $30+$ meter telescopes will likely reveal cold brown dwarfs with infrared spectral properties similar to dusty objects \citep{2019BAAS...51c.530D}. Since the relative flux change due to the Doppler boosting is $(\Delta F/F_{\rm em})_1\approx (\alpha-3)v_{\rm LOS}/c$ \citep{2020ApJ...905L..35R}, where $v_{\rm LOS}$ is a line-of-sight velocity (positive for receding objects), dust-enshrouded objects as well as pulsars are expected to exhibit a larger variability due to the beaming effect by at least a factor of a few in comparison with main-sequence stars.

In this study, we consider different negative values of a stellar spectral index in the infrared domain to quantify the relative flux variability due to the Doppler boosting. Since the S cluster hosts different types of objects, we focus on the enhanced variability with respect to main-sequence stars. 

The paper is structured as follows. In Section~\ref{sec_doppler_index}, we introduce the basic formalism of the continuum Doppler boosting and compare the relative flux variability among dust-enshrouded objects, pulsars and main-sequence stars. Subsequently, we evaluate the relative variability due to the Doppler boosting for four monitored dust-enshrouded objects in Section~\ref{sec_results} to demonstrate the importance of the spectral index for the actual objects in the S cluster. In Section~\ref{sec_discussion}, we discuss other internal processes for extended dust-enshrouded objects that can lead to significant continuum variations. We also consider the potential detection of longer beaming flares and dimming events. Finally, we conclude in Section~\ref{sec_conclusions}.  

\section{Doppler boosting of stellar light: effect of spectral index} \label{sec_doppler_index}

Stars orbiting the SMBH in the Galactic center can reach pericenter distances of a few 1000 gravitational radii ($r_{\rm g}$), reaching velocities of a few percent of the light speed. Two stars, S62 and S4714, are candidates to reach only a few 100 $r_{\rm g}$ with pericenter velocities up to $\sim 10\%$ of the light speed \citep{2020ApJ...899...50P}. As stars approach the SMBH, they are subject to time dilation, light aberration, and the frequency shift due to a stellar motion and a general relativistic redshift as well as a special relativistic transverse Doppler shift. All of these effects add up and lead to the Doppler boosting of stellar emission \citep{2006ApJ...639L..21Z,2007ApJ...670.1326Z,2020ApJ...905L..35R} that can make a star appear both fainter and brighter depending on the line-of-sight velocity.

The flux density of a star as detected by an observer on the Earth in a narrow frequency band $w(\nu)$ can be expressed as a power-law function of the ratio of observed and emitted frequencies $\nu_{\rm obs}$ and $\nu_{\rm em}$, respectively, as $F_{\rm obs}=F_{\rm em}(\nu_{\rm obs}/\nu_{\rm em})^{3-\alpha}$, where $F_{\rm em}=\int_0^{\infty}I_{\rm em}(\nu)w(\nu)\mathrm{d}\nu$ is the emitted flux density. 

\begin{figure}
    \centering
    \includegraphics[width=\columnwidth]{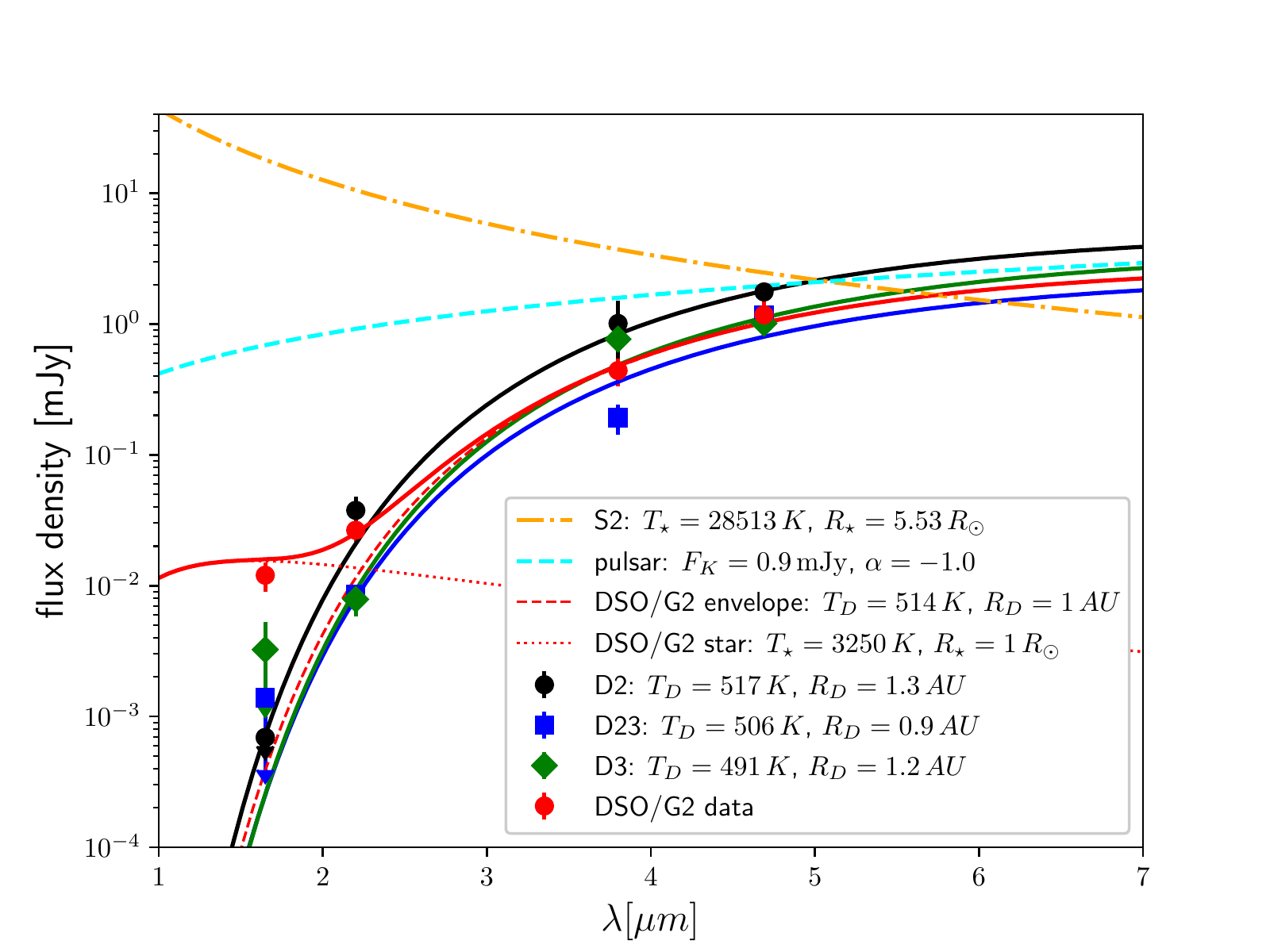}
    \caption{Representative spectral energy distributions (SED) of four dusty sources in the S cluster (D2, D23, D3, and DSO/G2) studied in \citet{2020A&A...634A..35P}, see the legend. We include measured flux densities in $H$, $K$, $L$, and $M$ bands as well as best-fit blackbody SEDs characterized by the temperature and the radius of the optically thick photosphere. For the DSO/G2 source, a two-component black-body fit is preferred consisting of a stellar black-body emission as well as a more extended dusty envelope. Hot main-sequence B-type stars in the S cluster are represented by the S2 star \citep[dot-dashed orange line; ][]{2017ApJ...847..120H}. A putative pulsar, whose energetic analogue is PSR B0540-69 \citep{2012A&A...544A.100M}, with the spectral slope $\alpha=-1.0$ is depicted by a cyan dashed line. Spectral slopes between $K$ and $L$ bands as well as $L$ and $M$ bands are listed in Table~\ref{tab_dusty_sources} for selected sources.}
    \label{fig_dusty_sources_S2}
\end{figure}

As shown by \citet{2020ApJ...905L..35R}, the relative flux variability due to $\mathcal{O}(v/c)$ terms (nonrelativistic Doppler beaming) for $v \ll c$ can be expressed as,
\begin{equation}
    \left(\frac{\Delta F}{F_{\rm em}} \right)_1\approx (\alpha-3)\frac{v_{\rm LOS}}{c}\,.
    \label{eq_doppler_nonrel}
\end{equation}
If we consider main-sequence B-type stars with $\alpha\sim 2$, in particular S2, the variability amplitude is $2.13\%$ and $2.07\%$ in $K$ ($2.2\,{\rm \mu m}$) and $L$ ($3.8\,{\rm \mu m}$) bands, respectively, see Table~\ref{tab_dusty_sources}. More exotic objects within the S cluster, namely monitored dusty objects \citep{2020A&A...634A..35P,2020Natur.577..337C} as well as putative pulsars \citep{2004ApJ...615..253P,2012ApJ...753..108W}, have a broad-band spectral energy distribution with the negative spectral index $\alpha<0$. Hence, the variability amplitude due to the Doppler boosting will be increased for them as the factor in Eq.~\eqref{eq_doppler_nonrel} is at least $\sim 3$ in absolute terms while for main-sequence stars with $\alpha\sim 2$ it is effectively equal to unity. As soon as more sensitive near- and mid-infrared detectors mounted at $30+$ telescopes are operational, it will be possible to probe the stellar initial mass function (IMF) towards the low-mass end where brown dwarfs of spectral type L and T are positioned \citep{2019BAAS...51c.530D}. Old brown dwarfs have effective temperatures $\lesssim 1000\,{\rm K}$ \citep{2000ARA&A..38..485B} and their SED slopes are thus comparable to dust-enshrouded objects and therefore the presented analysis is relevant for them as well. 

We take into account different negative values of the spectral index in the infrared $K$ and $L$ bands. First, we take into account the spectral energy distribution (SED) of dust-enshrouded objects D2, D23, D3, and DSO/G2 as analyzed by \citet{2020A&A...634A..35P}. Their SEDs are plotted in Fig.~\ref{fig_dusty_sources_S2}. In most cases, for D2, D23, and D3 sources, the broad-band SED can be described well as a one-component black-body spectrum with the effective temperature of $\sim 500\,{\rm K}$ and the radius of $\sim 1\,{\rm AU}$. For the DSO/G2 source, the two-component black-body fit is preferred since at the shortest wavelengths the stellar emission appears to dominate, see Fig.~\ref{fig_dusty_sources_S2} for the spectral decomposition and \citet{2020A&A...634A..35P} for a detailed discussion. The dominant contribution in $K$, $L$, and $M$ bands is due to the optically thick dusty envelope with the temperature of $T\sim 500\,{\rm K}$. The corresponding spectral indices between $K$ and $L$ bands as well as $L$ and $M$ bands -- $\alpha_{\rm KL}$ and $\alpha_{\rm LM}$ -- are listed in Table~\ref{tab_dusty_sources}. The mean values for the four studied dust-enshrouded objects are $\overline{\alpha}_{\rm KL}=-6.31 \pm 1.23$ and $\overline{\alpha}_{\rm LM}=-4.30 \pm 2.75$. We use these average values for a representative case of a dust-enshrouded object orbiting the SMBH on the S2 orbit, for which we obtain the variability amplitude of $18.16\%$ and $14.24\%$ in $K$ and $L$ bands, respectively. Hence, the variability due to the Doppler boosting is increased by a factor of $\sim 9$ and $\sim 7$ in $K$ and $L$ bands, respectively. If we define the characteristic half-maximum timescale $\tau_{1/2}$ between the epochs where the relative flux variability reaches half of the maximum value, we obtain $\tau_{1/2}=2.8$ years independent of the spectral index, which by definition affects only the amplitude of the relative flux variability.

Furthermore, the higher-order corrections to the Doppler boosting due to $\mathcal{O}(v^2/c^2)$ terms (gravitational redshift and the transverse Doppler shift) are also affected by the spectral index. If we consider the relative flux variability $\Delta F/F_{\rm em}$ up to $\mathcal{O}(v^2/c^2)$ terms, then we get \citep{2020ApJ...905L..35R},

\begin{align}
    \frac{\Delta F}{F_{\rm em}}&=\frac{F_{\rm obs}-F_{\rm em}}{F_{\rm em}}=\notag\\
    &=\left(\frac{1+\frac{v_{\rm LOS}}{c}}{1+(\phi_{\rm BH}-v^2/2)/c^2} \right)^{\alpha-3}-1\,,
    \label{eq_deltaF_Fem}
\end{align}
where $\phi_{\rm BH}=-GM_{\bullet}/r$ is the gravitational potential of the SMBH. Using Eq.~\eqref{eq_deltaF_Fem} and \eqref{eq_doppler_nonrel}, the higher-order corrections to the first-order Doppler boosting can be calculated as $(\Delta F/F_{\rm em})_2=\Delta F/F_{\rm em}-(\Delta F/F_{\rm em})_1$. In contrast to $\alpha=2$ considered by \citet{2020ApJ...905L..35R}, $(\Delta F/F_{\rm em})_2$ can reach positive values (negative magnitude change), which is unique for exotic objects within the S cluster, such as dusty objects, pulsars or brown dwarfs. The overall amplitude of the relative flux variability is also increased. For $\overline{\alpha}_{KL}=-6.31$ on the S2 orbit, we obtain $(\Delta F/F_{\rm em})_2=0.98\%$, while for $\overline{\alpha}_{LM}=-4.30$, we get $(\Delta F/F_{\rm em})_2=0.65\%$.  

\begin{figure}
    \centering
    \includegraphics[width=\columnwidth]{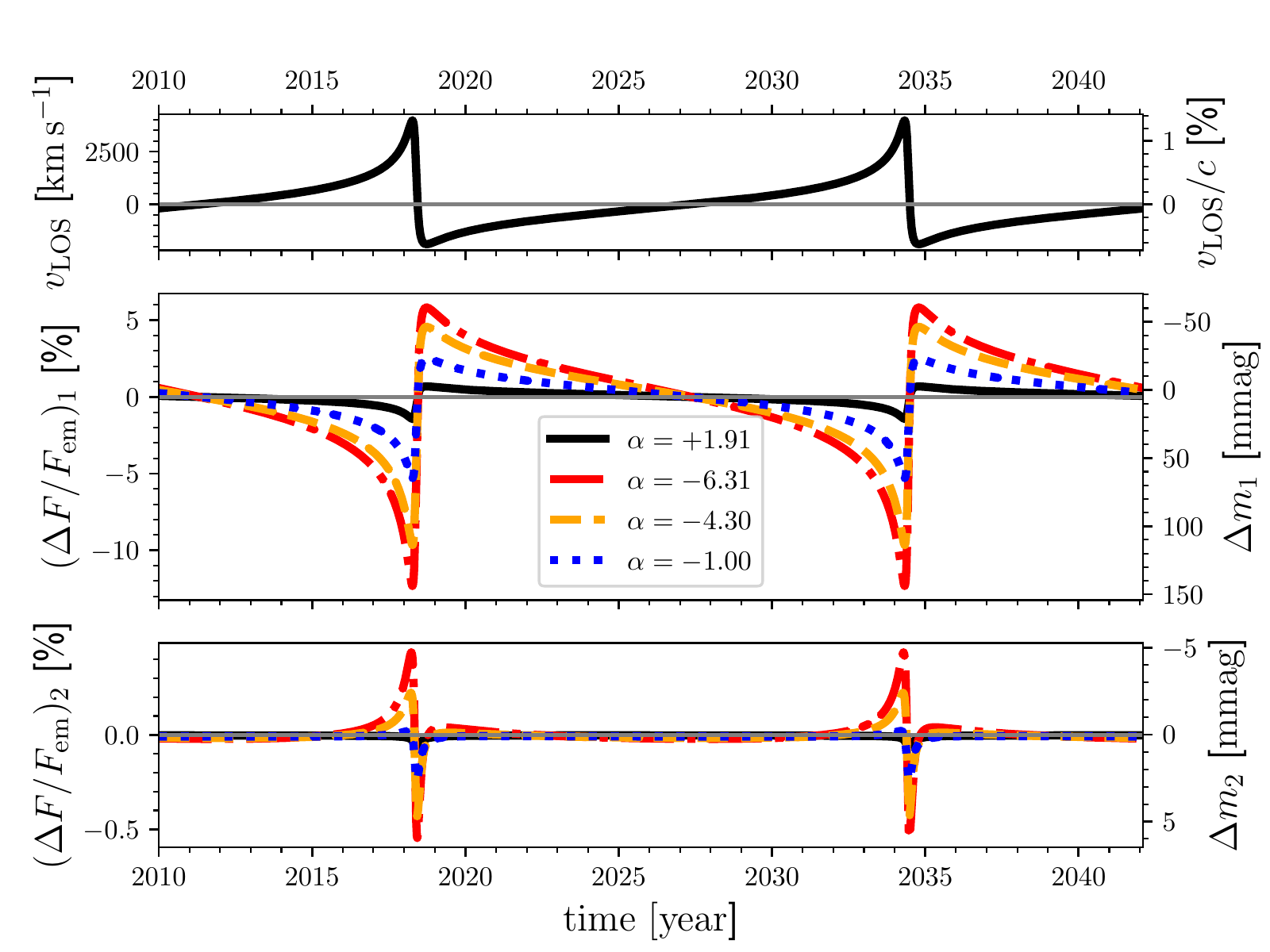}
    \caption{Effect of the near-infrared spectral slope on the Doppler-boosing effect of stellar light calculated for the star on the S2 orbit. {\bf Top panel:} A line-of-sight (LOS) velocity (in ${\rm km\,s^{-1}}$) as a function of time. The right $y$-axis shows the LOS velocity in the percent of the light speed. {\bf Middle panel:} The Doppler boosting due to the leading $\mathcal{O}(v/c)$ term (nonrelativistic) expressed as the relative flux change in percent. Different lines stand for different types of stellar objects present in the S cluster: B-type hot star (black solid line; $\alpha=+1.91$), a dust-enshrouded NIR-excess source observed in $K$ band (red dot-dashed line; $\overline{\alpha}_{\rm KL}=-6.31$), a dust-enshrouded NIR-excess source observed in $L$ band (orange dashed line; $\overline{\alpha}_{\rm LM}=-4.30$), and a pulsar (blue dotted line; $\alpha=-1.00$). The $y$-axis on the right shows the relative flux change in millimagnitudes (mmag). {\bf Bottom panel:} The relative flux change in percent due to $\mathcal{O}(v^2/c^2)$ terms (gravitational redshift and transverse Doppler shift). The $y$-axis on the right shows the relative change in millimagnitudes (mmag).}
    \label{fig_S2_alpha_comp}
\end{figure}

In Fig.~\ref{fig_S2_alpha_comp}, we show the line-of-sight velocity, $v_{\rm LOS}$ (top panel), the relative flux variability due to the Doppler boosting $(\Delta F/F_{\rm em})_1$ (middle panel) for different spectral indices, and the associated higher-order relative flux variability $(\Delta F/F_{\rm em})_2$ as a function of time for the exemplary orbit of the S2 star.

\begin{table*}[]
     \caption{Near-infrared spectral indices between $K$ and $L$ bands and $L$ and $M$ bands for the selected infrared-excess sources analyzed by \citet{2020A&A...634A..35P}. For comparison with main-sequence stars, we also include S2 star whose blackbody SED was calculated using the radius and the effective temperature from \citet{2017ApJ...847..120H}. We also modified the spectral index of S2 to demonstrate the impact on the beamed flux density for a dust-enshrouded object (slopes determined as mean values of four studied objects: D2, D23, D3, DSO/G2) as well as a pulsar. In a similar way, we also estimated the Doppler beaming for putative dust-enshrouded objects on S62- as well as S4714-like orbits. The amplitudes of relative flux variability due to $\mathcal{O}(v/c)$ terms, $\delta(\Delta F/F_{\rm em})_1$, as well as $\mathcal{O}(v^2/c^2)$ terms, $\delta(\Delta F/F_{\rm em})_2$, are calculated for both $K$ and $L$ infrared bands.}
    \centering
      \resizebox{\textwidth}{!}{
    \begin{tabular}{ccccccc}
    \hline
    \hline
Source &  $\alpha_{\rm KL}$ & $\alpha_{\rm LM}$ & $\delta(\Delta F/F_{\rm em})^{KL}_1$\,$[\%]$ & $\delta(\Delta F/F_{\rm em})^{LM}_1$\,$[\%]$ & $\delta(\Delta F/F_{\rm em})_2^{KL}$\,$[\%]$ & $\delta(\Delta F/F_{\rm em})_2^{LM}$\,$[\%]$ \\
    \hline
D2 &  -6.02 & -2.64 & $3.95$ & $2.47$ & $0.026$ & $0.011$  \\
D23 & -5.69 & -8.59 & $3.65$ & $4.86$ & $0.019$ & $0.033$ \\  
D3 & -8.38 &  -1.32 & $4.30$ & $1.63$ & $0.032$ & $0.0056$ \\
DSO/G2 & -5.15 & -4.65 & $16.79$ & $15.76$ & $0.66$ & $0.60$ \\
G1    & $(-9.07,1.17)$ & $(-5.50,0.89)$ & $18.33$ & $12.51$ & $1.02$ & $0.49$\\ 
S2     & +1.91  & +1.94 & $2.13$  & $2.07$ & $0.066$ & $0.064$ \\
pulsar (S2 orbit) & $-1.00$   & $-1.00$ & $7.80$ & $7.80$ & $0.26$ & $0.26$  \\
dust-enshrouded object (S2 orbit) & $-6.31$ & $-4.30$ & $18.16$ & $14.24$ & $0.98$ & $0.65$ \\
dust-enshrouded object (S62 orbit) & $-6.31$ & $-4.30$ & $60.27$ & $47.26$ & $12.25$   & $7.86$ \\
dust-enshrouded object (S4714 orbit) & $-6.31$ & $-4.30$ & $59.15$ & $46.38$ & $12.56$ & $7.61$\\
    \hline
    \end{tabular}
    }
    \label{tab_dusty_sources}
\end{table*}

Finally, we also consider a putative case of an energetic pulsar orbiting within the S cluster, which could mimic near-infrared excess sources in terms of the broad-band SED \citep{2017A&A...602A.121Z}. As an analogue, we adopt the energetics of PSR B0540-69 that has a prominent near-infrared emission \citep{2012A&A...544A.100M} with the $K$-band luminosity of $L_{\rm K}\approx 10^{34}\,{\rm erg\,s^{-1}}$. For the spectral index, we adopt $\alpha=-1$ that is typical of synchrotron-dominated young pulsars \citep{2012A&A...544A.100M}. The SED of such a putative pulsar is shown in Fig.~\ref{fig_dusty_sources_S2}. The amplitude of the relative flux variability is $7.80\%$, hence by about a factor of four larger than for the main-sequence star. The temporal evolution of $(\Delta F/F_{\rm em})_1$ is shown among other cases in Fig.~\ref{fig_S2_alpha_comp}. The amplitude of the relative variability due to higher-order corrections is also increased by a factor of $\sim 4$ to $\sim 0.26\%$. The flux density minimum of $(\Delta F/F_{\rm em})_2$ is at $+2.63\times 10^{-3}$ magnitudes, while the maximum at $-0.23 \times  10^{-3}$ magnitudes. The characteristic amplitudes of the Doppler boosting for both the first- and the second-order effect in $K$ and $L$ bands are summarized in Table~\ref{tab_dusty_sources} for the case of a main-sequence star, a dust-enshrouded object, as well as a pulsar assuming the same S2 orbit.

\section{Results}
\label{sec_results}

In this section, we describe the analysis of the relative flux variability due to Doppler beaming for monitored dust-enshrouded objects within the S cluster, specifically D2, D23, D3, and DSO/G2 objects. We adopt the infrared flux densities as determined by \citet{2020A&A...634A..35P}. The SEDs of these sources and the corresponding blackbody fits are shown in Fig.~\ref{fig_dusty_sources_S2}. All of these sources are on long-period orbits with periods of a few hundred years. Therefore the detection of a significant Doppler-boosted signal within the observational duration of a few years is unlikely with the upcoming monitoring or in archival data. However, we estimate the Doppler beaming effect for these sources to illustrate that the overall photometric amplitude of relative variations can be comparable to or even exceed the values expected for a short-period star S2 on a tight orbit. This is especially relevant in case such an infrared-excess source will be approaching or going through its pericenter during the future more sensitive photometric monitoring, such as previously G1 and DSO/G2 sources \citep{2014ApJ...796L...8W,2015ApJ...800..125V}.

Below we list relevant values of $(\Delta F/F_{\rm em})_1$, $(\Delta F/F_{\rm em})_2$ and their amplitudes for individual objects. 

\paragraph{D2} The source is on a mildly eccentric orbit with the eccentricity of $e\sim0.15$ and the semimajor axis of $a\sim 30\,{\rm mpc}$. The pericentre distance is $r_{\rm p}\sim 130\,000\,r_{\rm g}$ and the corresponding velocity is $v_{\rm p}\sim 900\,{\rm km\,s^{-1}}$. Therefore the resulting continuum variations are rather small, with the amplitudes $\delta(\Delta F/F_{\rm em})^{KL}_1=3.95\%$ and  $\delta(\Delta F/F_{\rm em})^{LM}_1=2.47\%$ for the first-order Doppler boosting. The amplitudes for the higher-order flux variations in the near-infrared $K$ and $L$ bands are $\delta(\Delta F/F_{\rm em})^{KL}_2=0.026\%$ and $\delta(\Delta F/F_{\rm em})^{LM}_2=0.011\%$, respectively. 

\paragraph{D23} This dusty source is the least eccentric from the selected sources, with $e=0.06\pm 0.01$. Orbiting the SMBH with the long period of $\sim 431$ years, its pericenter distance is as much as $\sim 212\,000\,r_{\rm g}$ with the corresponding velocity of $\sim 670\,{\rm km\,s^{-1}}$.  The first-order Doppler-boosting flux variations have the amplitude of $3.65\%$ and $4.86\%$ in $K$ and $L$ bands, respectively, while the relativistic corrections are two orders of magnitude smaller, $\delta(\Delta F/F_{\rm em})_2^{KL}=0.019\%$ and $\delta(\Delta F/F_{\rm em})_2^{LM}=0.033\%$. 

\paragraph{D3} This dusty source is on a mildly eccentric orbit with $e=0.24 \pm 0.02$ and the semimajor axis of $a=35.20 \pm 0.01\,{\rm mpc}$ (orbital period of $\sim 306$ years). Its smallest distance to Sgr~A* is $\sim 137\,000\,r_{\rm g}$ and the corresponding maximum velocity is $\sim 904\,{\rm km\,s^{-1}}$. Because of the measured steep slope between $K$ and $L$ bands, implying $\alpha\sim -8.4$, the amplitude of the first-order flux variability is $\delta(\Delta F/F_{\rm em})_1^{KL}=4.30\%$. Between $L$ and $M$ bands, the SED becomes flatter, which results in the smaller amplitude of $\delta(\Delta F/F_{\rm em})_1^{LM}=1.63\%$. The higher-order relativistic corrections are again at least two orders of magnitude smaller, with the amplitudes of $0.032\%$ and $0.0056\%$ in $K$ and $L$ bands, respectively.

\begin{figure}
    \centering
    \includegraphics[width=\columnwidth]{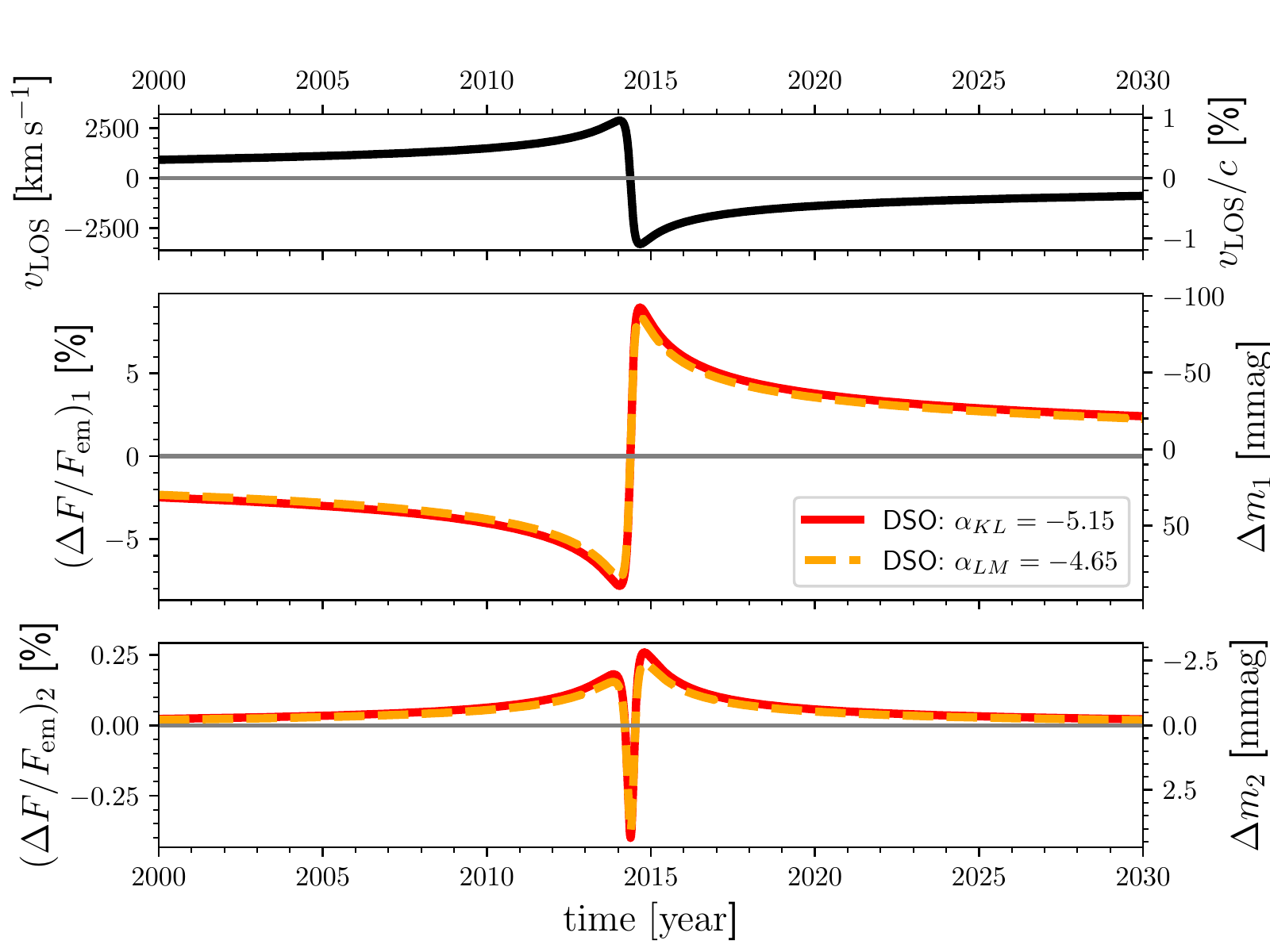}
    \caption{The same panels as in Fig.~\ref{fig_S2_alpha_comp} for the infrared-excess source DSO/G2. In the middle and the bottom panels, we plot the relative flux variability separately for the spectral index between $K$ and $L$ bands (red solid line) and for the spectral index between $L$ and $M$ bands (orange dashed line).}
    \label{fig_DSO}
\end{figure}

\paragraph{DSO/G2} The most eccentric source among the recently monitored dusty objects is DSO/G2 with $e=0.976$ \citep{2015ApJ...800..125V}. Having a semi-major axis of $a\sim 33\,{\rm mpc}$, it went through the pericenter of its orbit around $2014.39$, with the smallest distance of $r_{\rm p}\sim 4042\,r_{\rm g}$ and the maximum orbital velocity of $\sim 6600\,{\rm km\,s^{-1}}$. Given the high eccentricity and the resulting small pericenter distance, the amplitudes of the relative flux variability due to the leading-order Doppler beaming are the largest among the monitored infrared-excess sources, specifically $\delta(\Delta F/F_{\rm em})_1^{KL}=16.79\%$ ($0.182$ mag) and $\delta(\Delta F/F_{\rm em})_1^{LM}=15.76\%$ ($0.171$ mag). The amplitudes of the second-order relativistic variations are at the level of $\sim 1\%$, $\delta(\Delta F/F_{\rm em})^{KL}_2=0.66\%$ and $\delta(\Delta F/F_{\rm em})^{LM}_2=0.60\%$. The relative flux variations are depicted in Fig.~\ref{fig_DSO} as functions of time for the epochs close to the pericenter. The changes at the level of $\sim 0.2\,{\rm mag}$ are still challenging to detect in the crowded environment of the Galactic Center. The $L$-band light curve of the DSO/G2 as analyzed by \citet{2014ApJ...796L...8W} is constant within the observational uncertainties that reach as much as $\sim 0.5$ mag. The mean value of the flux slightly increased close to the pericenter and then dropped, however, this cannot be claimed as significant \citep{2014ApJ...796L...8W}.     

The compact dusty sources on highly eccentric orbits such as DSO/G2 provide the best prospects for detecting longer ``beaming flares'' or dimming events, i.e the phases when the star is expected to appear brighter or fainter for at least several months. While for D2, D23, and D3 sources we obtain the half-maximum timescales of $\tau_{1/2}=89$, $143$, and $111$ years, respectively, the DSO/G2 source has $\tau_{1/2}=3.55$ years, with the peak of $(\Delta F/F_{\rm em})_1 \sim 8.4\%-9.0\%$ ($-0.088$ and $-0.093$ mag, respectively) at $\sim 2014.67$ or $\sim 3.36$ months after reaching the pericenter.

We summarize the variability amplitudes $\delta (\Delta F/F_{\rm em})_1$ and $\delta (\Delta F/F_{\rm em})_2$ for the monitored dust-enshrouded objects in Table~\ref{tab_dusty_sources}.

\section{Discussion}
\label{sec_discussion}

We studied the effect of the infrared spectral index on the Doppler-beamed continuum emission of stars within the S cluster. The relative variability can be enhanced by a factor of a few up to an order of magnitude for negative values of the spectral index, which is exhibited by observed dust-enshrouded objects as well as putative young pulsars and older brown dwarfs.

However, especially the extended dust-enshrouded objects are prone to an order of magnitude variations due to tidal, irradiation, bow-shock, and ellipsoidal effects. These will prevent the detection of the clear Doppler-beamed signal. On the other hand, the expected Doppler-beaming variations can be subtracted to study the internal processes of these sources in more detail. We will discuss some of those other internal effects in the upcoming subsection. Subsequently, we also mention the possibility of detection of beaming flares at the position of Sgr~A* due to high-velocity dusty and other cold objects (brown dwarfs) as well as pulsars.

\subsection{Internal variations of dust-enshrouded objects}

There is an ongoing debate concerning the source of the continuum emission of dust-enshrouded objects, some of which are spatially extended. Currently, they appear to be internally heated due to the presence of a stellar core \citep{2017A&A...602A.121Z,2017ApJ...847...80W,2020A&A...634A..35P,2020Natur.577..337C}. However, the contribution of an external irradiation cannot be excluded. Below we discuss several processes that can likely contribute to the internal photometric variations of these objects, in particular those that are extended beyond their corresponding tidal (Hill) radii. Subtraction of the expected flux density changes due to Doppler beaming, provided that the orbital trajectory is well constrained, is relevant for the further assessment of these effects.

\begin{itemize}
\item[(a)] \textit{External irradiation.} Considering that there are $N_{\rm S} \sim 100$ stars within the S cluster of radius $R_{\rm S}\sim 1''\sim0.04\,{\rm pc}$, we obtain the mean stellar volume number density of $n_{\rm S}\sim 4\times 10^5\,{\rm pc^{-3}}$. The mean distance of an S star from any dusty object thus is $\overline{d}_{\rm S}=(n_{\rm S})^{-1/3}\sim 3\,000\,{\rm AU}$. Assuming that there is an embedded star inside the dusty object, its flux is greater than the flux from an S star by a factor
\begin{equation}
    \frac{F_{\star}}{F_{\rm S}}=\left(\frac{T_{\star}}{T_{\rm S}} \right)^4 \left(\frac{R_{\star}}{R_{\rm S}} \right)^2 \left(\frac{\overline{d}_{\rm S}}{R_{\rm DSO}} \right)^2\,,
    \label{eq_ratio_DSO_Sstar}
\end{equation}
where $F_{\star}$ is the flux of an embedded star, $T_{\star}$ and $R_{\star}$ are its temperature and radius, respectively, $F_{\rm S}$ is the flux of an external S star, $T_{\rm S}$ and $R_{\rm S}$ are its temperature and radius, respectively, and finally $R_{\rm DSO}$ is the characteristic DSO/G2 radius. Considering $T_{\star}=3250\,{\rm K}$ and $R_{\star}=1R_{\odot}$ from the two-component black-body fit, see Fig.~\ref{fig_dusty_sources_S2}, $T_{\rm S}\sim 30\,000\,{\rm K}$, $R_{\rm S}\sim 5\,R_{\odot}$ \citep{2017ApJ...847..120H}, and $R_{\rm DSO}\sim 1\,AU$ (Fig.~\ref{fig_dusty_sources_S2}), we obtain using Eq.~\eqref{eq_ratio_DSO_Sstar}, $F_{\star}/F_{\rm S}\sim 50$. The ratio $F_{\star}/F_{\rm S}$ is unity for the S star approach to the dusty object with the separation of $\overline{d}_{\rm S}\sim 424\,{\rm AU}$. The total bolometric magnitude change due to the external irradiation can approximately be estimated from the extreme approach of an S star within a few hundred AU, when its flux reprocessed by the gaseous-dusty envelope is comparable to the embedded star, $\Delta m\lesssim -2.5\log{2}\sim -0.8$ mag, which is greater than or comparable to the variations due to the Doppler-boosting effect.

\item[(b)] \textit{Ellipsoidal variations.} Due to tidal stretching and compression of the gaseous-dusty envelope, dusty objects are expected to have an ellipsoidal shape. As the object orbits around SMBH, the projected area seen by the observer changes, which leads to photometric variations. This effect was observed for a tidally distorted star orbiting the microquasar GRO J1655-40 \citep{2001ApJ...554.1290G}, where the amplitude of the optical and infrared continuum variations reaches $\sim 0.2-0.4\,{\rm mag}$.
\item[(c)] \textit{Bow-shock radiation.} Due to their supersonic motion at the pericenter, dust-enshrouded objects are expected to develop a bow shock, in which electrons are accelerated in the amplified magnetic field and emit broad-band synchrotron radiation. This can lead to the flux density enhancement close to the pericenter, which, however, depends on the ambient density as well as magnetic field. \citet{2020A&A...644A.105H} constrain the contribution of the non-thermal bow-shock emission to $\lesssim 0.1$ mag in near-infrared $L'$-band for the S2 star.
\item[(d)] \textit{Tidal truncation: Case of G1.}
A prominent and extended $L$-band emission source passed close to Sgr~A* around $2001.0$ \citep{2004A&A...417L..15C,2005A&A...439L...9C}. It was subsequently named as G1 and analyzed in detail by \citet{2017ApJ...847...80W} during its post-pericenter phase; see also \citet{2015ApJ...798..111P}. The source was more extended and resolved out before 2006 and afterwards got more compact and appeared as a point source, which was interpreted by the tidal truncation of its large gaseous-dusty envelope. Its $L$-band emission flux density also dropped by about two magnitudes \citep{2017ApJ...847...80W}.
\begin{figure}
    \centering
    \includegraphics[width=\columnwidth]{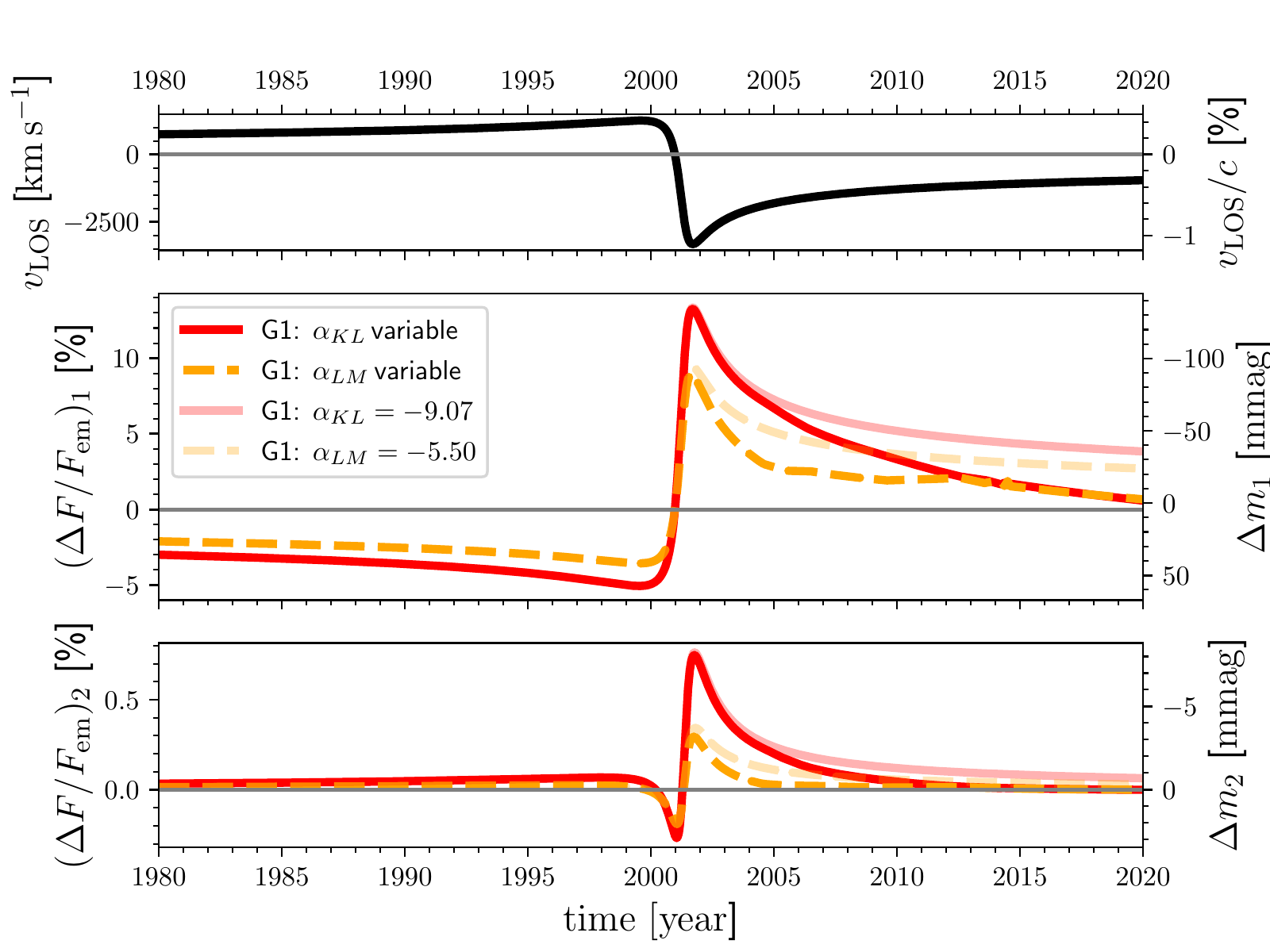}
    \caption{Effect of a variable spectral index on the Doppler beaming effect. The same panels as in Fig.~\ref{fig_S2_alpha_comp} for the infrared-excess source G1. In the middle and the bottom panels, we plot the relative flux variability separately for the spectral index between $K$ and $L$ bands (red solid line) and for the spectral index between $L$ and $M$ bands (orange dashed line). In particular, we consider increasing spectral indices as indicated by the observations of \citet{2017ApJ...847...80W}. For comparison, we also show, in fainter colours, the case with fixed spectral indices in $K$ and $L$ bands.}
    \label{fig_G1object}
\end{figure}
The first-order and second-order Doppler beaming as a function of time is depicted in Fig.~\ref{fig_G1object}. We compare two cases: the Doppler beaming for a gradually increasing spectral index (bright red and orange lines) as indicated by the analysis of \cite{2017ApJ...847...80W} and the case that assumes fixed spectral indices (fainter red and orange lines). The main effect that we see is that the evolving spectral slope in the post-pericenter phase effectively shortens the duration of the beaming ``flare'' (see also Section~\ref{subsection_flares}), while the amplitude is not significantly affected. For the leading-term Doppler boosting effect, we obtain the amplitudes $\delta(\Delta F/F_{\rm em})_1^{KL}=18.33\%$ ($\sim 0.192$ magnitudes) and $\delta(\Delta F/F_{\rm em})_1^{LM}=12.51\%$ ($\sim 0.132$ magnitudes). The higher-order relativistic corrections have the overall amplitudes of $\delta(\Delta F/F_{\rm em})_2^{KL}=1.02\%$ ($\sim 0.011$ magnitudes) and $\delta(\Delta F/F_{\rm em})_2^{LM}=0.49\%$ ($\sim 0.005$ magnitudes). Since there is no $L$-band light curve of G1 close to its pericenter passage available and the $L$-band flux density drops by nearly two magnitudes due to structural changes in the post-pericenter phase of the orbit, the Doppler-boosting effect cannot be practically traced for this object. On the other hand, the subtraction of nearly $\sim 0.1-0.2\,{\rm mag}$ variations due to the Doppler boosting is of relevance for studying internal changes of the source close to the pericenter passage. 
\end{itemize}

For future monitoring, compact dusty sources, whose intrinsic $K$ and $L$-band flux densities are rather stable within at least $0.1$ magnitude, such as the DSO/G2 object, appear suitable for the detection of the Doppler beaming effect, especially due to the fact that the overall amplitude of the boosting effect can be larger by an order of magnitude in comparison with B-type S stars. Cold brown dwarfs orbiting Sgr~A* that are even more compact and are expected to be photometrically stable are even more suitable for a detailed monitoring of the Doppler beaming. Observations by future 30$+$ telescopes equipped with adaptive optics in near- and mid-infrared bands should be sensitive enough to detect faint objects orbiting Sgr~A* down to the brown-dwarf level \citep{2019BAAS...51c.530D}.

\subsection{Beaming flares and dimming of $\sim 0.1-1$ mag close to Sgr~A*}
\label{subsection_flares}

Faint dust-enshrouded objects, pulsars, and brown dwarfs on highly eccentric orbits observed in near-infrared bands are expected to cause brightening events or beaming ``flares'' with an increase of $\sim 0.1-1$ magnitudes and lasting for a few days to years, depending on the pericenter distance. Having these properties, the events would be unique and could be distinguished from the near-infrared flares of Sgr~A* that last approximately one hour and occur a few times per day \citep{2020arXiv201109582W}.

Specifically, focusing on $K$-band observations, for a dust-enshrouded object on the S2 orbit we get the peak relative flux of $5.83\%$ ($\Delta m=-0.062$ magnitudes) with the timescale of $\tau_{1/2}=2.79$ years. For a pulsar on the same orbit, we get the relative flux maximum of $2.50\%$ ($\Delta m=-0.027$ magnitudes) with the same timescale. For the DSO/G2 object, the relative flux density increases up to $8.97\%$ ($\Delta m=-0.093$ magnitudes) with the half-maximum timescale of $\tau_{1/2}=3.55$ years.


\begin{figure*}
    \centering
    \includegraphics[width=\columnwidth]{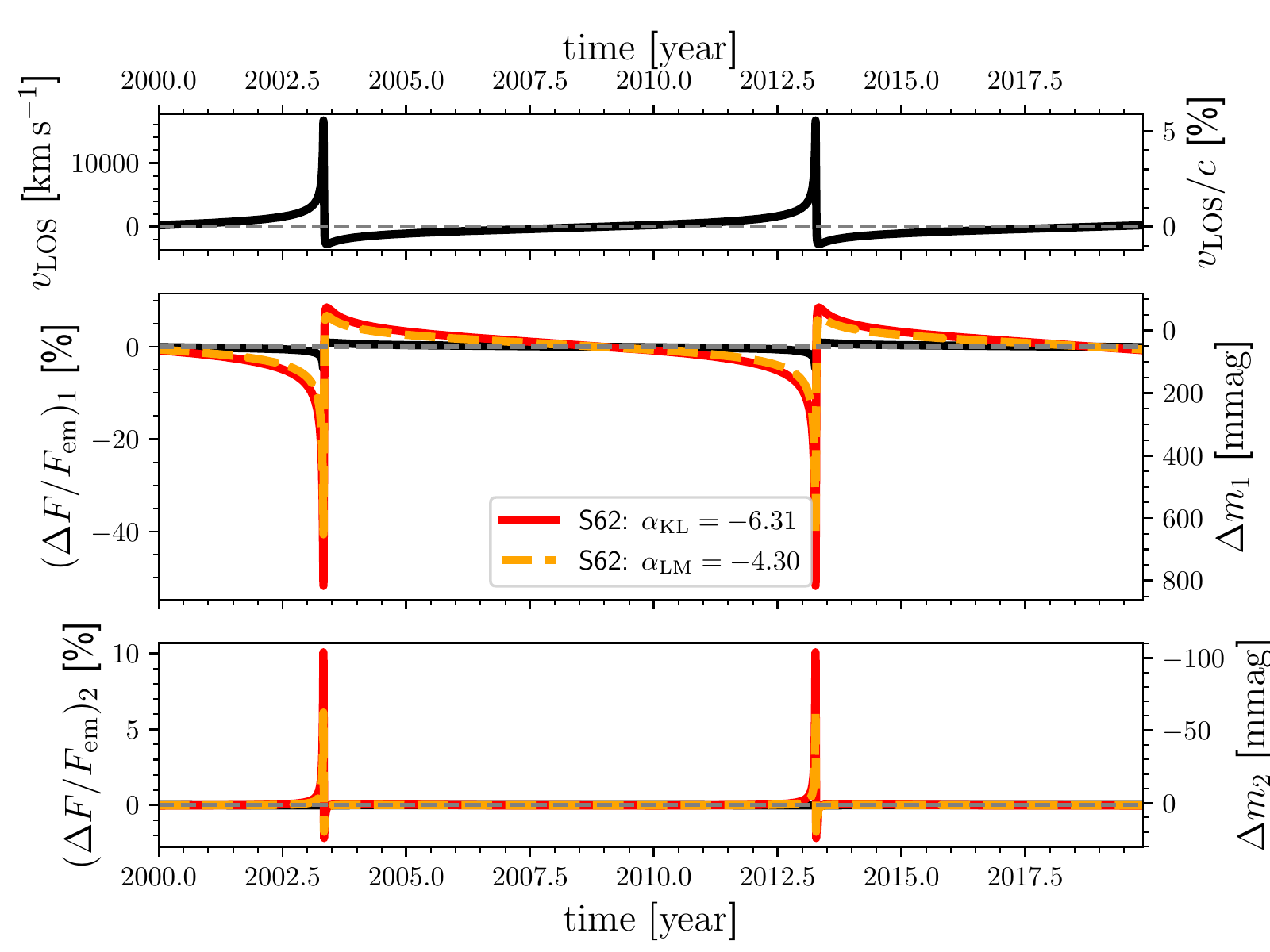}
     \includegraphics[width=\columnwidth]{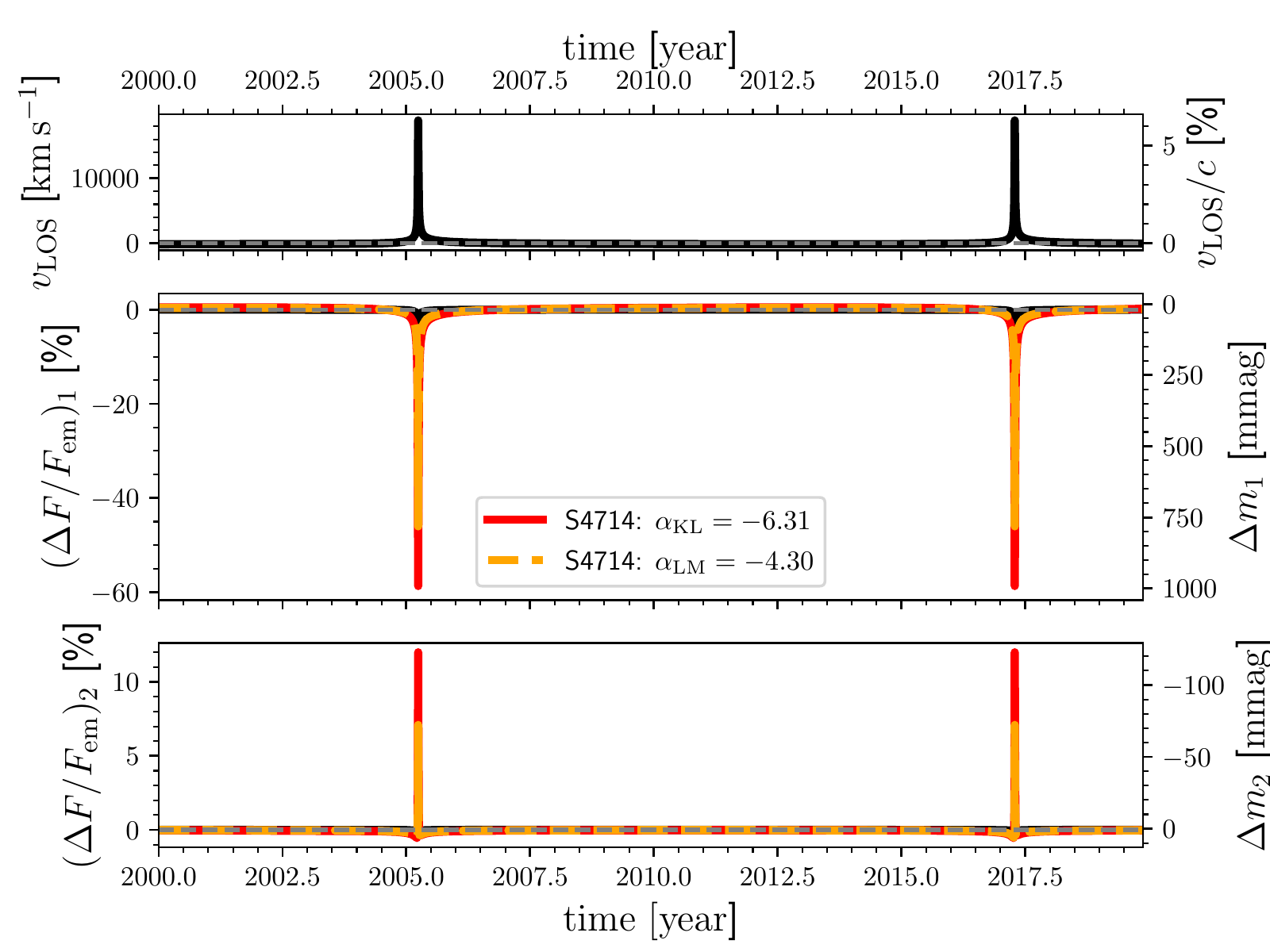}
    \caption{The same panels as in Fig.~\ref{fig_S2_alpha_comp} for a dust-enshrouded object on an S62-like orbit (left panel) and on an S4714-like orbit (right panel). In the middle and the bottom panels, we plot the relative flux variability separately for the spectral index between $K$ and $L$ bands (red solid line) and for the spectral index between $L$ and $M$ bands (orange dashed line).}
    \label{fig_S62_S4714}
\end{figure*}

Even more prominent Doppler beaming can occur at the pericenter distances one order of magnitude less than for the S2 star. Recently, \citet{2020ApJ...889...61P} and \citet{2020ApJ...899...50P} reported the detection of faint S stars, some of which, in particular S62 and S4714, can approach Sgr~A* at the distance as small as $\sim 439\,r_{\rm g}$ and $\sim 312\,r_{\rm g}$, respectively. \citet{2020ApJ...905L..35R} predicted the variability of $\sim 6\%$ for both S62 and S4714. In case a dust-enshrouded object with the $K$-band spectral index of $\overline{\alpha}_{\rm KL}=-6.31$ would be moving on an orbit comparable to S62 and S4714, the variability could reach as much as $\sim 60\%$ for S62-like orbit and $\sim 59\%$ for S4714-like orbit, see Table~\ref{tab_dusty_sources} for the summary of amplitudes and Fig.~\ref{fig_S62_S4714} for the temporal profiles of $(\Delta F/F_{\rm em})_1$ and $(\Delta F/F_{\rm em})_2$ for both stars.

As we can see in light curves of the relative variability in Fig.~\ref{fig_S62_S4714}, both for S62 and S4714 orbits the Doppler dimming by a few $\sim 0.1$ magnitudes would be more prominent than the Doppler brightening. Especially for faint dusty objects, brown dwarfs or pulsars on similar orbits, there would be an epoch of dimming or potentially a complete disappearance of the object in case its infrared flux would be at the sensitivity limit. For S62-like orbit, the minimum relative flux is $\sim -52\%$ ($\Delta m\sim 0.8$ magnitudes) and the half-minimum timescales is $\tau_{1/2}\sim 0.036$ years or $\sim 13$ days. In case of S4714, the dimming minimum is $\sim 59\%$ ($\Delta m\sim 1$ magnitude) and the half-minimum timescale is $\tau_{1/2}=4.75$ days.

\section{Conclusions}
\label{sec_conclusions}

Motivated by the occurrence of colder dust-enshrouded objects among main-sequence B-type stars, we explored the effect of the infrared spectral index on the Doppler beaming effect within the S cluster. For three dust-enshrouded objects on mildly eccentric orbits, D2, D23, and D3, we found the relative variability amplitudes of $3.95\%$, $3.65\%$, and $4.30\%$ in $K$-band, respectively. The most prominent variability is for highly eccentric dusty objects DSO/G2 and G1 with the the $K$-band amplitudes of $16.79\%$ and $18.33\%$, respectively, where the latter object has a gradually increasing spectral index. In general, dust-enshrouded objects with steep spectral indices are expected to exhibit an enhanced Doppler beaming effect by as much as an order of magnitude in comparison with B-type S stars. Putative synchrotron-powered pulsars would have a stronger Doppler-beaming variability by a factor of about four.

Therefore, the best prospects for detecting a Doppler beaming effect within the S cluster with future sensitive instruments, such as Extremely Large Telescope, are for compact dusty objects, brown dwarfs, and pulsars that are expected to have stable intrinsic continuum flux densities. For extended dusty objects, such as G1, variations due to tidal, ellipsoidal, bow-shock, and irradiation effects are generally larger. The subtraction of the expected Doppler-beaming variations will help to better comprehend their internal physics, especially close to the pericenter passage.

In case the Doppler-beamed emission is detected for objects with well-determined orbits, the temporal profiles of the first-  and second-order Doppler boosting could be utilized to constrain the spectral index, and hence, the nature of the objects. Moreover, with the expected detection of fainter cold compact objects within the S cluster (dust-enshrouded objects, brown dwarfs) using the upcoming sensitive infrared photometry, we expect that the Doppler boosting will lead to episodic beaming ``flares'' or dimming by a few $0.1$ magnitudes on the timescale of a year within $\sim 15$ milliarcseconds of Sgr~A*.

\acknowledgments
\section*{acknowledgments}

The author thanks the referee for constructive comments that helped to improve the manuscript. The project was partially supported by the Polish Funding Agency National Science Centre, project 2017/26/A/ST9/00756 (MAESTRO 9). MZ also acknowledges the NAWA financial support under
the agreement PPN/WYM/2019/1/00064 to perform a
three-month exchange stay at the Charles University
in Prague and the Astronomical Institute of the Czech
Academy of Sciences in Prague. A part of the presented research was performed within the Polish-Czech mobility program ((MŠMT 8J20PL037 and NAWA PPN/BCZ/2019/1/00069).  







\end{document}